\documentclass[twocolumn,showpacs,preprintnumbers,amsmath,amssymb]{revtex4}

\usepackage{graphicx}
\usepackage{dcolumn}
\usepackage{bm}

\usepackage{amssymb}
\usepackage{amsmath}

\begin{document}

\title{Cosmological model from the holographic equipartition law \\ with a modified R\'{e}nyi entropy}

\author{Nobuyoshi {\sc Komatsu}}  \altaffiliation{E-mail: komatsu@se.kanazawa-u.ac.jp} 

\affiliation{Department of Mechanical Systems Engineering, Kanazawa University, 
                          Kakuma-machi, Kanazawa, Ishikawa 920-1192, Japan }

\begin{abstract}
Cosmological equations were recently derived by Padmanabhan from the expansion of cosmic space due to the difference between the degrees of freedom on the surface and in the bulk in a region of space. In this study, a modified R\'{e}nyi entropy is applied to Padmanabhan's `holographic equipartition law', by regarding the Bekenstein--Hawking entropy as a nonextensive Tsallis entropy and using a logarithmic formula of the original R\'{e}nyi entropy. Consequently, the acceleration equation including an extra driving term (such as a time-varying cosmological term) can be derived in a homogeneous, isotropic, and spatially flat universe. When a specific condition is mathematically satisfied, the extra driving term is found to be constant-like as if it is a cosmological constant. Interestingly, the order of the constant-like term is naturally consistent with the order of the cosmological constant measured by observations, because the specific condition constrains the value of the constant-like term. 

\end{abstract}

\pacs{98.80.-k, 98.80.Es, 95.30.Tg}

\maketitle

\section{Introduction} 
\label{Introduction}
To explain the accelerated expansion of the late universe, $\Lambda$CDM (lambda cold dark matter) models assume a cosmological constant $\Lambda$ related to an additional energy component called dark energy \cite{PERL1998_Riess1998,Riess2007SN1,Planck2015}. However, $\Lambda$ measured by observations is many orders of magnitude smaller than the theoretical value estimated by quantum field theory \cite{Weinberg1989}. This discrepancy is the so-called cosmological constant problem. To resolve this theoretical difficulty, numerous cosmological models have been proposed \cite{Weinberg1,Bamba1}, such as CCDM (creation of CDM) models \cite{Prigogine_1988-1989,Lima-Others1996-2016,Lima_1992-2016} and $\Lambda (t)$CDM models, which assume a time-varying cosmological term $\Lambda (t)$ \cite{Freese-Mimoso_2015,Sola_2009-2015,Sola_2013_Review,Sola_2015_2,Sola_2015L14,Sola_2016_1,Valent2015,LimaSola_2013a,LimaSola_2015-2016,EPJC}. In CCDM models, a constant term is obtained from a dissipation process based on gravitationally induced particle creation \cite{Prigogine_1988-1989,Lima-Others1996-2016,Lima_1992-2016}, while in $\Lambda(t)$CDM models, a constant term is obtained from an integral constant of the renormalization group equation for the vacuum energy density \cite{Sola_2013_Review}. 

For these models, thermodynamic scenarios have attracted attention, in which the Bekenstein--Hawking entropy (which is proportional to the surface area of the event horizon) \cite{Bekenstein1,Hawking1} and the holographic principle (which refers to the information of the bulk stored on the horizon) \cite{Hooft-Bousso} play important roles. For example, using the holographic principle it has been proposed that gravity is itself an entropic force derived from changes in the Bekenstein--Hawking entropy \cite{Jacob1995,Padma1,Verlinde1,Padma2010}. Based on this concept, the cosmological equations have been extensively examined in a homogeneous and isotropic universe \cite{Sheykhi1,Sadjadi1}, although the cosmological constant has not been discussed. In an alternative treatment, Easson \textit{et al.} \cite{Easson12} proposed an entropic cosmology that assumes the usually neglected surface terms on the horizon of the universe \cite{Koivisto-Costa1,Lepe1,Basilakos1-Sola_2014a,Koma4,Koma5,Koma6,Koma7,Koma8,Koma9,Gohar_2015ab}. In entropic cosmology, an extra driving term to explain the accelerated expansion is derived from entropic forces on the horizon of the universe. An area entropy (the Bekenstein--Hawking entropy), a volume entropy (the Tsallis--Cirto entropy) \cite{Tsallis2012}, a quartic entropy \cite{Koma5,Koma6}, and a general form of entropy \cite{Koma9,Gohar_2015ab} have been applied to entropic cosmology. However, the entropic-force term (related to dark energy and $\Lambda$) is generally considered to be a tuning parameter, which makes it difficult to include in discussion of the cosmological constant problem.

Padmanabhan recently provided a new insight into the origin of spacetime dynamics using another thermodynamic scenario called the `holographic equipartition law' \cite{Padma2012A}. Based on the holographic equipartition law with Bekenstein--Hawking entropy, cosmological equations in a flat universe can be derived from the expansion of cosmic space due to the difference between the degrees of freedom on the surface and in the bulk \cite{Padma2012A}. The emergence of the cosmological equations (i.e., the Friedmann and acceleration equations) has been examined from various viewpoints, such as a non-flat universe and quantum-corrected entropy \cite{Padma2012,Cai2012-Tu2013,Padma2014-2015,ZLWang2015,Tu2015}. However, dark energy and $\Lambda$ have not yet been discussed fundamentally, though they have been considered in several studies \cite{Padma2014-2015,ZLWang2015,Tu2015}. This is likely because dark energy and $\Lambda$ (which are related to an extra driving term) have been assumed to explain the accelerating universe. If the extra driving term is naturally derived from the holographic equipartition law, it is possible to study various cosmological models based on holographic equipartition law. However, such a driving term should not be derived using Bekenstein--Hawking entropy.

Self-gravitating systems exhibit peculiar features due to long-range interacting potentials. Therefore, the R\'{e}nyi entropy \cite{Ren1} and the Tsallis entropy \cite{Tsa0} can also be used for astrophysical problems \cite{Tsallis2012,Tsa1,Czinner1,Czinner2,Plas1-Tsallis2001,Chava21-Liu,NonExtensive,Koma2-3,Nunes_2015b,Czinner2016}. Bir\'{o} and Czinner \cite{Czinner1} have recently suggested a novel type of R\'{e}nyi entropy on a black-hole horizon, in which the Bekenstein--Hawking entropy is considered to be a nonextensive Tsallis entropy \cite{Czinner1,Czinner2}.

In this paper, the modified R\'{e}nyi entropy is applied to the holographic equipartition law in a homogeneous, isotropic, and spatially flat universe. It is expected that an extra driving term can be derived from the holographic equipartition law with the modified R\'{e}nyi entropy. Therefore, the present study should help in developing cosmological models based on the holographic equipartition law. In addition, the extra driving term is expected to be constant-like under specific conditions. The specific condition and constant-like term may provide new insights into the cosmological constant problem. (The present study focuses on the derivation of the extra driving term and the specific condition for the constant-like term. Accordingly, the inflation of the early universe and density perturbations related to structure formations are not examined.)

The remainder of the article is organized as follows.
In Sec.\ \ref{LCDM models}, $\Lambda (t)$CDM models are briefly reviewed for a typical formulation of the cosmological equations. In Sec.\ \ref{Entropy on the horizon}, entropies on the horizon are discussed. The Bekenstein--Hawking entropy is reviewed in Sec.\ \ref{Bekenstein-Hawking entropy}, while a modified R\'{e}nyi is introduced in Sec.\ \ref{Renyi entropy}. The holographic equipartition law is discussed in Sec.\ \ref{Holographic equipartition}. In Sec.\ \ref{Renyi entropy and holographic equipartition law}, the modified R\'{e}nyi entropy is applied to the holographic equipartition law, to derive the acceleration equation that includes an extra driving term. The extra driving term is then discussed in Sec.\ \ref{Discussions}, focusing on the specific condition required for obtaining a constant-like term. Finally, in Sec.\ \ref{Conclusions}, the conclusions of the study are presented.

\section{$\Lambda(t)$CDM models} 
\label{LCDM models}
Cosmological equations derived from the holographic equipartition law are expected to be similar to those for $\Lambda (t)$CDM models \cite{Freese-Mimoso_2015,Sola_2009-2015,Sola_2013_Review,Sola_2015_2,Sola_2015L14,Sola_2016_1,Valent2015,LimaSola_2013a,LimaSola_2015-2016,EPJC} in a nondissipative universe.
Therefore, in this section, the $\Lambda(t)$CDM model is reviewed, to discuss a typical formulation of the cosmological equations.

A homogeneous, isotropic, and spatially flat universe is considered, and the scale factor $a(t)$ is examined at time $t$ in the Friedmann--Lema\^{i}tre--Robertson--Walker metric  \cite{Koma6,Koma9}. In the $\Lambda(t)$CDM model, the Friedmann equation is given as 
\begin{equation}
  \left(  \frac{ \dot{a}(t) }{ a(t) } \right)^2  =  H(t)^2     =  \frac{ 8\pi G }{ 3 } \rho (t) +  \frac{\Lambda(t)}{3}   , 
\label{eq:L(t)_FRW01}
\end{equation}
and the acceleration equation is  
\begin{equation}
  \frac{ \ddot{a}(t) }{ a(t) }   =  \dot{H}(t) + H(t)^{2}   
                                          =  -  \frac{ 4\pi G }{ 3 } \left ( \rho (t) + \frac{3p(t)}{c^2}  \right ) +  \frac{\Lambda(t)}{3} ,  
\label{eq:L(t)_FRW02}
\end{equation}
where the Hubble parameter $H(t)$ is defined by
\begin{equation}
   H(t) \equiv   \frac{ da/dt }{a(t)} =   \frac{ \dot{a}(t) } {a(t)}  .
\label{eq:Hubble}
\end{equation}
$G$, $c$, $\rho(t)$, and $p(t)$ are the gravitational constant, the speed of light, the mass density of cosmological fluids, and the pressure of cosmological fluids, respectively \cite{Koma6,Koma9}, and $\Lambda(t)$ is a time-varying cosmological term. Based on Eqs.\ (\ref{eq:L(t)_FRW01}) and (\ref{eq:L(t)_FRW02}), the continuity equation \cite{Koma6} is given by
\begin{equation}
       \dot{\rho}(t) + 3  \frac{\dot{a}(t)}{a(t)} \left (  \rho (t) + \frac{p(t)}{c^2}  \right )    =  - \frac{\dot{\Lambda}(t)}{8 \pi G}        .
\label{eq:drho_L(t)}
\end{equation}
The right-hand side of this continuity equation is usually non-zero, except for the case $\Lambda(t)$=$\Lambda$. Accordingly, the $\Lambda(t)$CDM model can be interpreted as a kind of energy exchange cosmology in which the transfer of energy between two fluids is assumed \cite{Barrow22-2015}. When $\Lambda(t)$=$\Lambda$, the Friedmann, acceleration, and continuity equations are identical to those for the standard $\Lambda$CDM model. In this paper, cosmological models based on the holographic equipartition law are assumed to be a particular case of $\Lambda (t)$CDM models, although the theoretical backgrounds are different.

\section{Entropy on the horizon} 
\label{Entropy on the horizon}

In the holographic equipartition law, the horizon of the universe is assumed to have an associated entropy \cite{Padma2012A}. The Bekenstein--Hawking entropy is generally used, replacing the horizon of a black hole by the horizon of the universe. In Sec.\ \ref{Bekenstein-Hawking entropy}, the Bekenstein--Hawking entropy is briefly reviewed. In Sec.\ \ref{Renyi entropy}, a novel type of R\'{e}nyi entropy proposed by Bir\'{o} and Czinner \cite{Czinner1} is introduced for the entropy on the horizon of the universe.

\subsection{The Bekenstein--Hawking entropy} 
\label{Bekenstein-Hawking entropy}

The Bekenstein--Hawking entropy $S_{\rm{BH}}$ \cite{Bekenstein1} is given as
\begin{equation}
 S_{\rm{BH}}  = \frac{ k_{B} c^3 }{  \hbar G }  \frac{A_{H}}{4}   ,
\label{eq:SBH}
\end{equation}
where $k_{B}$ and $\hbar$ are the Boltzmann constant and the reduced Planck constant, respectively \cite{Koma4,Koma5}. The reduced Planck constant is defined by $\hbar \equiv h/(2 \pi)$, where $h$ is the Planck constant. $A_{H}$ is the surface area of the sphere with the Hubble horizon (radius) $r_{H}$, given by
\begin{equation}
     r_{H} = \frac{c}{H}   .
\label{eq:rH}
\end{equation}
Substituting $A_{H}=4 \pi r_{H}^2 $ into Eq.\ (\ref{eq:SBH}) and using Eq.\ (\ref{eq:rH}), we obtain \cite{Koma4,Koma5} 
\begin{equation}
S_{\rm{BH}}  = \frac{ k_{B} c^3 }{  \hbar G }   \frac{A_{H}}{4}       
                  =  \left ( \frac{ \pi k_{B} c^5 }{ \hbar G } \right )  \frac{1}{H^2}  
                  =    \frac{K}{H^2}    , 
\label{eq:SBH2}      
\end{equation}
or equivalently
\begin{equation}
S_{\rm{BH}} =   \frac{K}{H^2}   =\frac{  \pi  k_{B}  c^2 }{ L_{p}^{2} H^2}   , 
\label{eq:SBH2_2}      
\end{equation}
where $K$ is a positive constant given by
\begin{equation}
  K =  \frac{  \pi  k_{B}  c^5 }{ \hbar G } = \frac{  \pi  k_{B}  c^2 }{ L_{p}^{2} } , 
\label{eq:K-def}
\end{equation}
and $L_{p}$ is the Planck length, written as
\begin{equation}
  L_{p} = \sqrt{ \frac{\hbar G} { c^{3} } } .
\label{eq:Lp}
\end{equation}
As shown in Eqs.\ (\ref{eq:SBH}) and (\ref{eq:SBH2_2}), the Bekenstein--Hawking entropy $S_{\rm{BH}}$ on the Hubble horizon is proportional to $H^{-2}$ (and $A_{H}$) and is related to the Planck length.

In this paper, a spatially flat universe, $k = 0$, is considered, where $k$ is a curvature constant. For a spatially non-flat universe ($k \neq 0$), the apparent horizon given by $r_{A} = c/\sqrt{H^2 + (k /a^2) }$ is used instead of the Hubble horizon, see, e.g., Ref.\ \cite{Cai2012-Tu2013}.

\subsection{Modified R\'{e}nyi entropy} 
\label{Renyi entropy}

Nonextensive entropies for black hole thermodynamics have been extensively investigated \cite{Tsallis2012,Czinner1,Czinner2,NonExtensive}. Recently, Bir\'{o} and Czinner \cite{Czinner1} suggested a novel type of R\'{e}nyi entropy on black-hole horizons, by regarding the Bekenstein--Hawking entropy as a nonextensive Tsallis entropy and using a logarithmic formula. In the present study, the modified R\'{e}nyi entropy is used for the entropy on the Hubble horizon. Therefore, the modified R\'{e}nyi entropy \cite{Czinner1,Czinner2} is briefly reviewed here.

In nonextensive thermodynamics, the Tsallis entropy $S_{T}$ for a set of $W$ discrete states is defined as 
\begin{equation}
S_{T}  =  k_{B} \frac{  1- \sum_{i=1}^{W} p_{i}^{q} }{   q-1  }    ,
\label{eq:ST}
\end{equation}
where $p_{i}$ is a probability distribution and $q$ is the so-called nonextensive parameter \cite{Tsallis2012,Tsa1}. In addition, the original R\'{e}nyi entropy \cite{Tsallis2012,Ren1} is defined as 
\begin{equation}
S_{R}^{\rm{org}}  =  k_{B} \frac{ \ln \sum_{i=1}^{W} p_{i}^{q} }{   1-q  }  =  \frac{1}{1-q} \ln [1+ (1-q) S_{T} ] .
\label{eq:SR0}
\end{equation}
When $q  = 1$, both $S_{T}$ and $S_{R}^{\rm{org}}$ recover the Boltzmann--Gibbs entropy given by
\begin{equation}
S_{\textrm{BG}} =  - k_{B}  \sum_{i=1}^{W} p_{i} \ln p_{i}    .
\label{eq:S-BG}
\end{equation}
The original R\'{e}nyi entropy \cite{Tsallis2012,Ren1} can be written as
\begin{equation}
S_{R}^{\rm{org}}  = \frac{1}{1-q} \ln [1+ (1-q) S_{T} ] = \frac{1}{\lambda} \ln [1+ \lambda S_{T} ]   ,  
\label{eq:SR}
\end{equation}
where, for simplicity, $1-q$ has been replaced by $\lambda$: 
\begin{equation}
   \lambda = 1-q          .
\label{eq:lambda}
\end{equation}
In this paper, $\lambda$ is considered to be non-negative, as discussed later.

A novel type of R\'{e}nyi entropy has been proposed and examined in Refs.\ \cite{Czinner1,Czinner2}, in which not only is the logarithmic formula of the original R\'{e}nyi entropy used but the Bekenstein--Hawking entropy $S_{\rm{BH}}$ is considered to be a nonextensive Tsallis entropy $S_{T}$. Using Eq.\ (\ref{eq:SR}) and replacing $S_{T}$ by $S_{\rm{BH}}$, we obtain a modified R\'{e}nyi entropy $S_{R}$ \cite{Czinner2} given by 
\begin{equation}
S_{R} = \frac{1}{\lambda} \ln [1+ \lambda S_{\rm{BH}} ]   . 
\label{eq:SR-SBH}
\end{equation}
When $\lambda =0$ (i.e., $q  = 1$), $S_{R}$ becomes $S_{\rm{BH}}$. We call $S_{R}$ the modified R\'{e}nyi entropy. 
In the present study, the modified R\'{e}nyi entropy is applied to the holographic equipartition law.
To this end, the holographic equipartition law is reviewed in the next section.

Note that the physical origin of the modified R\'{e}nyi entropy is likely unclear at the present time. However, it is worthwhile to examine cosmological models based on the holographic equipartition law from various viewpoints. Therefore, in this paper, the modified R\'{e}nyi entropy is applied to the holographic equipartition law.

\section{Holographic equipartition law} 
\label{Holographic equipartition}

Padmanabhan \cite{Padma2012A} derived the Friedmann and acceleration equations in a flat universe from the expansion of cosmic space due to the difference between the degrees of freedom on the surface and in the bulk in a region of space. In this section, the Padmanabhan idea called the `holographic equipartition law' is reviewed, based on his work \cite{Padma2012A} and other related works \cite{Padma2012,Cai2012-Tu2013,Padma2014-2015,ZLWang2015,Tu2015}. (The surface terms assumed in entropic cosmology \cite{Easson12} are not considered in the holographic equipartition law. In Ref.\ \cite{Tu2015}, the holographic equipartition law is examined, including the surface terms.)

In an infinitesimal interval $dt$ of cosmic time, the increase $dV$ of the cosmic volume can be expressed as 
\begin{equation}
     \frac{dV}{dt}  =  L_{p}^{2} (N_{\rm{sur}} - \epsilon N_{\rm{bulk}} ) \times c      , 
\label{dVdt_N-N}
\end{equation}
where $N_{\rm{sur}}$ is the number of degrees of freedom on the spherical surface of Hubble radius $r_{H}$ and $N_{\rm{bulk}}$ is the number of degrees of freedom in the bulk \cite{Padma2012A}. $L_{p}$ is the Planck length given by Eq.\ (\ref{eq:Lp}) and $\epsilon$ is a parameter discussed later.
Equation\ (\ref{dVdt_N-N}) represents the so-called holographic equipartition law. Note that the right-hand side of Eq.\ (\ref{dVdt_N-N}) includes $c$, because $c$ is not set to be $1$ in the present paper. Using $r_{H}= c/H$ given by Eq.\ (\ref{eq:rH}), the Hubble volume $V$ can be written as
\begin{equation}
V = \frac{4 \pi}{3} r_{H}^{3} =  \frac{4 \pi}{3} \left ( \frac{c}{H} \right )^{3}   .
\label{eq:V}
\end{equation}
The number of degrees of freedom in the bulk is assumed to obey the equipartition law of energy \cite{Padma2012A}: 
\begin{equation}
  N_{\rm{bulk}} = \frac{|E|}{ \frac{1}{2} k_{B} T}     , 
\label{N_bulk}
\end{equation}
where the Komar energy $|E|$ contained inside the Hubble volume $V$ is given by 
\begin{equation}
|E| =  |( \rho c^2 + 3p)| V  = - \epsilon ( \rho c^2 + 3p) V  ,
\label{Komar}
\end{equation}
and $\epsilon$ is a parameter defined as  \cite{Padma2012A,Padma2012} 
 \begin{equation}
        \epsilon \equiv     
 \begin{cases}
              +1  & (\rho c^2 + 3p <0  )  ,  \\ 
              -1  & (\rho c^2 + 3p >0   )  .  \\
\end{cases}
\label{epsilon}
\end{equation}
From this definition, $|E|$ is confirmed to be nonnegative. Note that $\rho c^2 + 3p <0$ corresponds to an accelerating universe (e.g., dark energy dominated universe), while $\rho c^2 + 3p >0$ corresponds to a decelerating universe (e.g., matter and radiation dominated universe).  The temperature $T$ on the horizon is written as
\begin{equation}
 T = \frac{ \hbar H}{   2 \pi  k_{B}  }   .
\label{eq:T0}
\end{equation}
This temperature is used for calculating $N_{\rm{bulk}}$ from Eq.\ (\ref{N_bulk}) \cite{C1}. In contrast, the number of degrees of freedom on the spherical surface is given by  
\begin{equation}
  N_{\rm{sur}} = \frac{4 S_{H} }{k_{B}}       , 
\label{N_sur}
\end{equation}
where $S_{H}$ is the entropy on the Hubble horizon. When $S_{H} = S_{\rm{BH}}$, Eq.\ (\ref{N_sur}) is equivalent to that in Ref.\ \cite{Padma2012A}. In the next section, $S_{H}$ is replaced by the modified R\'{e}nyi entropy $S_{R}$ given by Eq.\ (\ref{eq:SR-SBH}).

We now derive the cosmological equations from the holographic equipartition law. To this end, we first calculate the left-hand side of Eq.\ (\ref{dVdt_N-N}). Substituting Eq.\ (\ref{eq:V}) into Eq.\ (\ref{dVdt_N-N}), the left-hand side of Eq.\ (\ref{dVdt_N-N}) is given by
\begin{equation}
     \frac{dV}{dt}  =      \frac{d}{dt} \left ( \frac{4 \pi}{3} \left ( \frac{c}{H} \right )^{3}  \right )   =  -4 \pi c^{3}   \left (  \frac{ \dot{H} }{H^{4} } \right )  . 
\label{dVdt_right}
\end{equation}
In this calculation, $r$ has been set to be $r_{H}$ before the time derivative is calculated \cite{Padma2012A}. Next, we calculate $N_{\rm{bulk}}$ included in the right-hand side of Eq.\ (\ref{dVdt_N-N}). According to Ref.\ \cite{Padma2012A}, $\rho c^2 + 3p <0$ is selected and, therefore, $\epsilon = +1$ from Eq.\ (\ref{epsilon}). Note that the following discussion is not affected even if $\rho c^2 + 3p >0$ is selected \cite{Padma2012}. Substituting Eqs.\ (\ref{Komar}) and (\ref{eq:T0}) into Eq.\ (\ref{N_bulk}), using Eq.\ (\ref{eq:V}), and rearranging, we obtain 
\begin{align}
  N_{\rm{bulk}}  &= \frac{|E|}{ \frac{1}{2} k_{B} T}  =    \frac{ - ( \rho c^2 + 3p) V  }{ \frac{1}{2} k_{B}  \left (  \frac{ \hbar H}{   2 \pi  k_{B}  }  \right )  }  
                         =    \frac{ -  ( \rho c^2 + 3p) \frac{4 \pi}{3} \left ( \frac{c}{H} \right )^{3}   }{ \frac{1}{2} k_{B}  \left ( \frac{ \hbar H}{   2 \pi  k_{B}  }  \right )   }   \notag \\
                      &=   -  \frac{ (4 \pi)^{2} c^{5}  }{ 3 \hbar } \left (  \rho  + \frac{3p}{c^{2}}  \right )  \frac{1}{  H^{4}   }      .
\label{N_bulk_cal}
\end{align}
In addition, substituting $\epsilon = +1$ and Eqs.\ (\ref{eq:Lp}), (\ref{N_sur}), (\ref{dVdt_right}), and (\ref{N_bulk_cal}) into Eq.\ (\ref{dVdt_N-N}) gives 
\begin{equation}
 \frac{ -4 \pi c^{3}   \dot{H} }{H^{4} }   = \frac{\hbar G} { c^{3} }  \left [ \frac{4 S_{H} }{k_{B}} + \frac{ (4 \pi)^{2} c^{5}  }{ 3 \hbar } \left (  \rho  + \frac{3p}{c^{2}}  \right )  \frac{1}{  H^{4}   }  \right ] \times c     ,
\label{dVdt_N-N_cal1}
\end{equation}
and solving this with regard to $\dot{H}$, we have
\begin{align}
  \dot{H}  &=  - \frac{ H^{4} }{ 4 \pi c^{3} }   \frac{\hbar G} { c^{3} }  \left [ \frac{4 S_{H} }{k_{B}} +  \frac{ (4 \pi )^{2} c^{5}  }{ 3 \hbar } \left (  \rho  + \frac{3p}{c^{2}}  \right )  \frac{1}{  H^{4}   }  \right ]  \times c   \notag \\
               &=   -  \frac{ 4 \pi G }{ 3} \left (  \rho  + \frac{3p}{c^{2}}  \right )  - \left ( \frac{ \hbar G }{  \pi k_{B} c^{5} } \right )  S_{H} H^{4}                                                                                                             \notag \\
               &=   -  \frac{ 4 \pi G }{ 3} \left (  \rho  + \frac{3p}{c^{2}}  \right )  - \frac{ S_{H} H^{4} }{K}                                                                                                    , 
\label{dVdt_N-N_cal2}
\end{align}
where $K$ is given by Eq.\ (\ref{eq:K-def}).
As shown in Eq.\ (\ref{eq:L(t)_FRW02}), $\ddot{a}/ a$ is written as $ \ddot{a}/ a   =  \dot{H} + H^{2}$.
Substituting Eq.\ (\ref{dVdt_N-N_cal2}) into this equation, we obtain 
\begin{align}
  \frac{ \ddot{a} }{ a }       &=  \dot{H} + H^{2}              \notag \\
                                      &=   -  \frac{ 4 \pi G }{ 3} \left (  \rho  + \frac{3p}{c^{2}}  \right )  - \frac{ S_{H} H^{4} }{K}      + H^{2}  \notag \\
                                      &=   -  \frac{ 4 \pi G }{ 3} \left (  \rho  + \frac{3p}{c^{2}}  \right ) + H^{2}  \left (  1 - \frac{ S_{H} H^{2} }{K}  \right )     .
\label{N-N_FRW02_SH}
\end{align}
The above equation is the acceleration equation derived from the holographic equipartition law, where $S_{H}$ is the entropy on the Hubble horizon. Interestingly, the second term on the right-hand side, i.e., $H^{2}(1-S_{H}H^{2}/K)$, appears to be an extra driving term.
 
When the Bekenstein--Hawking entropy is used, i.e., $S_{H} = S_{\rm{BH}}$, the second term $H^{2}(1-S_{H}H^{2}/K)$ is zero because $S_{\rm{BH}} = K/H^{2}$ given by Eq.\ (\ref{eq:SBH2}). Consequently, the acceleration equation can be written as 
\begin{equation}
  \frac{ \ddot{a} }{ a }      = -  \frac{ 4 \pi G }{ 3} \left (  \rho  + \frac{3p}{c^{2}}  \right )   .
\label{N-N_FRW02_S_BH}
\end{equation}
The obtained cosmological equation is considered to be a particular case of $\Lambda(t)$CDM models, as mentioned in Sec.\ \ref{LCDM models}. Accordingly, Eq.\ (\ref{N-N_FRW02_S_BH}) indicates that $\Lambda(t)$ in Eq.\ (\ref{eq:L(t)_FRW02}) is zero. Substituting $\Lambda(t)=0$ into Eqs.\ (\ref{eq:L(t)_FRW01}) and (\ref{eq:drho_L(t)}), we have the Friedmann and continuity equations given by  
\begin{equation}
  H^2     =  \frac{ 8\pi G }{ 3 } \rho  ,  
\label{eq:FRW01_L=0}
\end{equation}
\begin{equation}
       \dot{\rho} + 3  \frac{\dot{a}}{a} \left (  \rho + \frac{p}{c^2}  \right )    =  0 .
\label{eq:drho_L=0}
\end{equation}
The three equations (i.e., the Friedmann, acceleration, and continuity equations) agree with those in Ref.\ \cite{Padma2012A}.

In the above, the second term on the right-hand side of Eq.\ (\ref{N-N_FRW02_SH}) is zero because the Bekenstein--Hawking entropy $S_{\rm{BH}}$ is used for the entropy $S_{H}$ on the horizon. However, if a different $S_{H}$ is assumed, the second term (corresponding to an extra driving term) is expected to be non-zero. In the next section, we examine this expectation, applying the modified R\'{e}nyi entropy $S_{R}$ to the holographic equipartition law. Note that a similar driving term has been studied, e.g., in Ref.\ \cite{Tu2015}, using quantum corrected entropy.

\section{Acceleration equation from the holographic equipartition law with a modified R\'{e}nyi entropy} 
\label{Renyi entropy and holographic equipartition law}

In this section, the modified R\'{e}nyi entropy $S_{R}$ is applied to the holographic equipartition law, instead of the Bekenstein--Hawking entropy. The modified R\'{e}nyi entropy given by Eq.\ (\ref{eq:SR-SBH}) can be written as
\begin{equation}
S_{R} = \frac{1}{\lambda} \ln [1+ \lambda S_{\rm{BH}} ]   , 
\label{eq:SR-SBH_2}
\end{equation}
where $S_{R}$ becomes $S_{\rm{BH}}$ when $\lambda =0$.
The acceleration equation derived from the holographic equipartition law given by Eq.\ (\ref{N-N_FRW02_SH}) is 
\begin{equation}
  \frac{ \ddot{a} }{ a }       
                                     =   -  \frac{ 4 \pi G }{ 3} \left (  \rho  + \frac{3p}{c^{2}}  \right ) + H^{2}  \left (  1 - \frac{ S_{H} H^{2} }{K}  \right )     , 
\label{N-N_FRW02_SH_2}
\end{equation}
where $S_{H}$ is the entropy on the Hubble horizon and $K$ given by Eq.\ (\ref{eq:K-def}) is written as
\begin{equation}
  K =  \frac{  \pi  k_{B}  c^5 }{ \hbar G } = \frac{  \pi  k_{B}  c^2 }{ L_{p}^{2} } .
\label{eq:K-def_2}
\end{equation}

Regarding $S_{R}$ as $S_{H}$ and substituting Eq.\ (\ref{eq:SR-SBH_2}) into Eq.\ (\ref{N-N_FRW02_SH_2}), we have
\begin{align}
  \frac{ \ddot{a} }{ a }       
                                   & =   -  \frac{ 4 \pi G }{ 3} \left (  \rho  + \frac{3p}{c^{2}}  \right ) + H^{2}  \left (  1 - \frac{ \frac{1}{\lambda} \ln [1+ \lambda S_{\rm{BH}} ]    H^{2} }{K}  \right )      \notag \\
                                   & =   -  \frac{ 4 \pi G }{ 3} \left (  \rho  + \frac{3p}{c^{2}}  \right ) + f(H)    ,         
\label{N-N_FRW02_SR0}
\end{align}
where, using $S_{\rm{BH}}= K/H^{2}$ given by Eq.\ (\ref{eq:SBH2}), the extra driving term $f(H)$ is written as 
\begin{equation}
 f(H)  = H^{2}  \left (  1 - \frac{  \ln [1+ (\lambda K /H^{2}) ]  }{  \lambda K /H^{2} }  \right )      ,
\label{f(H)_1}
\end{equation}
and $\lambda$ is a non-negative constant, i.e., $\lambda \geq 0$. The acceleration equation given by Eq.\ (\ref{N-N_FRW02_SR0}) can be derived from the holographic equipartition law with the modified R\'{e}nyi entropy. When $\lambda =0$ (i.e., $S_{R} = S_{\rm{BH}}$), $f(H)$ is zero, as discussed in the previous section. However, $f(H)$ is expected to be non-zero when $\lambda > 0$. Consequently, $f(H)$ given by Eq.\ (\ref{f(H)_1}) is not a simple power series of $H$ but a logarithmic formula.

\begin{figure} [t]  
\begin{minipage}{0.495\textwidth}
\begin{center}
\scalebox{0.32}{\includegraphics{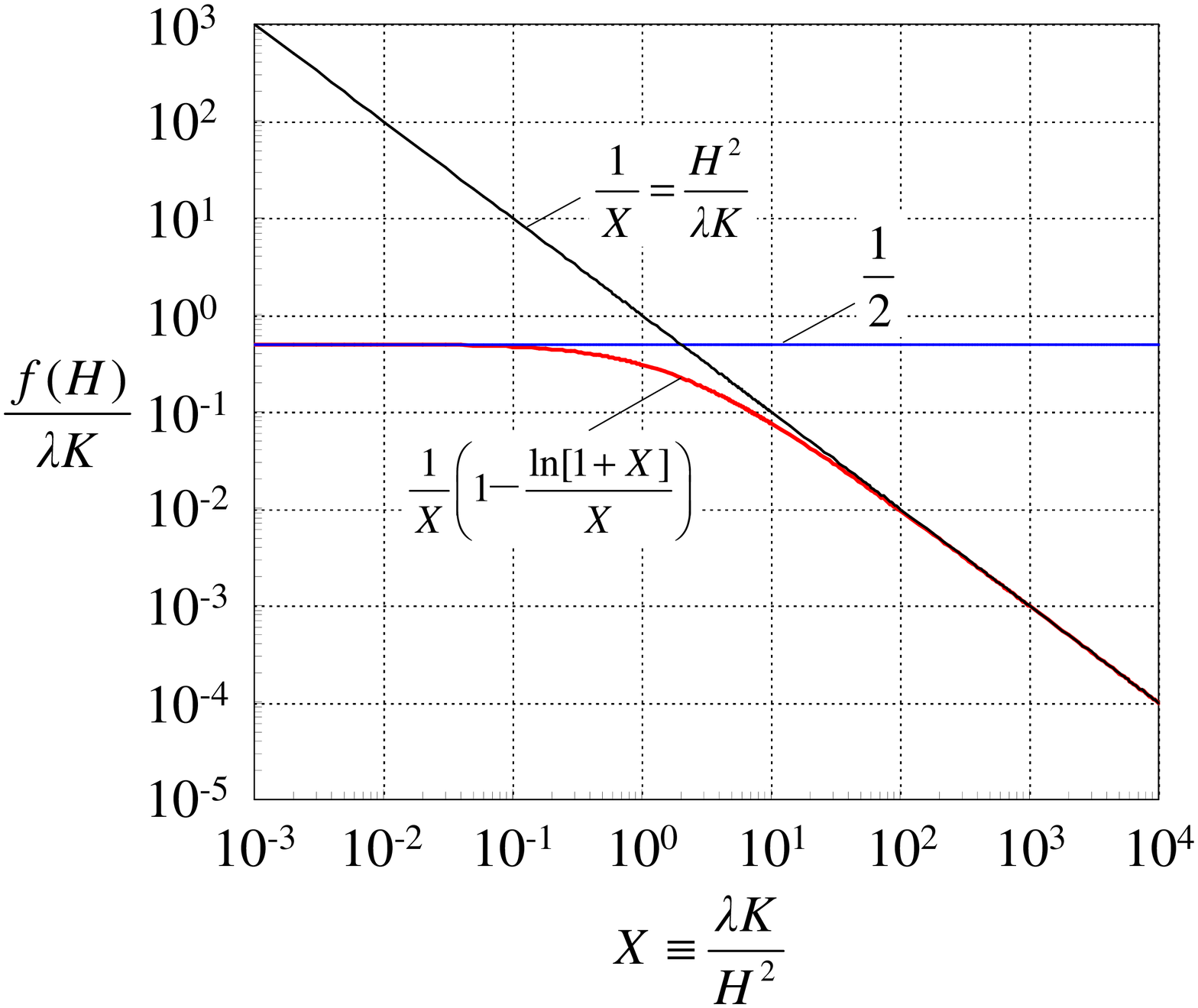}}
\end{center}
\end{minipage}
\caption{ (Color online). Properties of the normalized extra-driving term $f(H)/(\lambda K)$.  The horizontal axis is $X \equiv \lambda K/H^{2}$. The logarithmic formula given by Eq.\ (\ref{f(H)LK_1}) is plotted. The two approximation curves for $1/2$ and $1/X$ given by Eq.\ (\ref{f(H)LK_approx}) are also shown. }
\label{Fig-f-X}
\end{figure}

From Eq.\ (\ref{f(H)_1}), $f(H)$ can be written as
\begin{align}
 f (H)       &=  H^{2}  \left (  1 - \frac{  \ln [1+ (\lambda K /H^{2}) ]  }{  \lambda K /H^{2} }  \right )                                                      \notag \\
               &= \frac{ \lambda K}{ \lambda K /H^{2} }   \left (  1 - \frac{  \ln [1+ (\lambda K /H^{2}) ]  }{  \lambda K /H^{2} }  \right )         .     
\label{f(H)_002}
\end{align}
The positive dimensionless parameter $X$ is defined as 
\begin{equation}
  X  \equiv  \frac{   \lambda K  }{  H^{2}  }     \quad  (> 0) , 
\label{eq:X-def}
\end{equation}
where $\lambda$ is considered to be positive, i.e., $\lambda > 0$. Dividing Eq.\ (\ref{f(H)_002}) by $\lambda K$, and applying Eq.\ (\ref{eq:X-def}), we have the normalized extra-driving term given by
\begin{equation}
   \frac{ f (H) }{\lambda K}  = \frac{1} { X  }   \left (  1 - \frac{  \ln [1+ X ]  }{  X }  \right )     .
\label{f(H)LK_1}
\end{equation}
To observe the properties of $f(H)$, the normalized term is plotted in Fig.\ \ref{Fig-f-X}. Consequently, $f(H)/(\lambda K)$ gradually approaches $1/2$ with decreasing $X$, while it approaches a common curve with increasing $X$. Therefore, we focus on two specific conditions, $ X \leq 1$ and $X \gg 1 $. Under the two conditions, the approximate formulas are given mathematically by
\begin{equation}
\ln [1+ X ]  \approx X - \frac{X^2}{2} + \cdots    \quad    (X  \leq 1 )      ,
\label{ln(1+X)}
\end{equation}
\begin{equation}
\frac{\ln [1+ X ]}{X} \approx 0      \quad    (X  \gg 1 )      .
\label{ln(1+X)/X}
\end{equation}
High-order terms have been neglected in Eq.\ (\ref{ln(1+X)}).
Substituting Eqs.\ (\ref{ln(1+X)}) and (\ref{ln(1+X)/X}), respectively, into Eq.\ (\ref{f(H)LK_1}), we have 
 \begin{equation}
        \frac{ f (H) }{\lambda K} \approx     
 \begin{cases}
              \frac{1}{2}  & ( X \leq 1     )  ,  \\ 
              \frac{1}{X}  & ( X \gg 1  )  .   \\
\end{cases}
\label{f(H)LK_approx}
\end{equation}
From Eqs.\ (\ref{eq:X-def}) and (\ref{f(H)LK_approx}), $f(H)$ can be written as
 \begin{equation}
     f(H)  \approx     
 \begin{cases}
                \frac{\lambda K}{2}        & ( \frac{   \lambda K  }{  H^{2}  } \leq 1     )  ,  \\ 
                H^{2}                             & ( \frac{   \lambda K  }{  H^{2}  } \gg 1  )     .  \\
\end{cases}
\label{f(H)_approx}
\end{equation}
The constant-like and $H^{2}$-like terms are respectively obtained from each condition, as shown in Eq.\ (\ref{f(H)_approx}). The constant-like term, $\lambda K /2$, can be interpreted as a kind of cosmological constant, although the specific condition $\lambda K  /  H^{2}  \leq 1$ is required. In the next section, we discuss the constant-like term and the specific condition.

As a matter of fact, the extra driving term (which behaves like effective dark energy) can be derived even if dark energy is not assumed, i.e., even when $p=0$. Alternatively, the modified R\'{e}nyi entropy is assumed in the present paper. In this sense, the original holographic equipartition law is modified. This modification is expected to provide new insights into cosmological models based on the holographic equipartition law.

As shown in Eq.\ (\ref{f(H)_approx}), $H^{2}$-like terms are derived under a specific condition. 
However, the driving term is not likely to be related to the inflation of the early universe, because higher exponents terms such as $H^{4}$ terms should be required for the inflation  \cite{Koma9}. In $\Lambda (t)$CDM models, the higher exponents terms have been closely examined in Ref.\ \cite{Sola_2015_2}. In entropic cosmology, quantum corrected entropy is introduced for the higher exponents terms, see, e.g., Ref.\ \cite{Tu2015} and the second paper of Ref.\ \cite{Easson12}.

\section{Constant-like term under a specific condition}
\label{Discussions}

The acceleration equation can be derived from the holographic equipartition law with the modified R\'{e}nyi entropy, as examined in the previous section. Interestingly, an extra driving term included in the equation is found to be constant-like under a specific condition. In this section, we discuss the constant-like term under the specific condition. To this end, the present cosmological model is reviewed, assuming that it is a particular case of $\Lambda(t)$CDM models. Substituting $\Lambda(t)/3 = f(H)$ into Eqs.\ (\ref{eq:L(t)_FRW01}), (\ref{eq:L(t)_FRW02}), and (\ref{eq:drho_L(t)}), we obtain the Friedmann, acceleration, and continuity equations written as
\begin{equation}
    H^2     =  \frac{ 8\pi G }{ 3 } \rho  +  f(H) ,  
\label{N-N_FRW01_SR1}
\end{equation}
\begin{equation}
  \frac{ \ddot{a} }{ a }    =   -  \frac{ 4 \pi G }{ 3} \left (  \rho  + \frac{3p}{c^{2}}  \right ) + f(H)    ,         
\label{N-N_FRW02_SR1}
\end{equation}
\begin{equation}
       \dot{\rho} + 3  \frac{\dot{a}}{a} \left (  \rho + \frac{p}{c^2}  \right )    =  - \frac{3 \dot{f}(H)}{8 \pi G}        ,
\label{drho_SR1}
\end{equation}
where the extra driving term $f(H)$ is given by
\begin{equation}
 f(H)  = H^{2}  \left (  1 - \frac{  \ln [1+ (\lambda K /H^{2}) ]  }{  \lambda K /H^{2} }  \right )  ,   
\label{f(H)_2}
\end{equation}
and $K$ is 
\begin{equation}
  K =  \frac{  \pi  k_{B}  c^5 }{ \hbar G } = \frac{  \pi  k_{B}  c^2 }{ L_{p}^{2} } .
\label{eq:K-def_3}
\end{equation}
As shown in Eq.\ (\ref{drho_SR1}), the right-hand side of the continuity equation is not zero generally. This non-zero right-hand side may be interpreted as the interchange of energy between the bulk (the universe) and the surface (the horizon of the universe) \cite{Lepe1,Koma9}. (The previous works of Sol\`{a} \textit{et al.} \cite{Sola_2015L14,Sola_2016_1} imply that the value of the non-zero right-hand side of the continuity equation should be small in $\Lambda (t)$CDM models \cite{Koma9}.) In the following, the right-hand side of the continuity equation is zero, because $f(H)$ is considered to be constant under a specific condition. General solutions for the present model, which includes a logarithmic term, are separately discussed in Appendix \ref{Solutions for the present model and evolutions of the universe}.

This paper focuses on the constant-like term under a specific condition. From Eq.\ (\ref{f(H)_approx}), the constant-like term and the specific condition can be written as
 \begin{equation}
     f(H)  \approx  \frac{\lambda K}{2} =  \left ( \frac{\lambda K}{2H_{0}^{2}}  \right )  H_{0}^{2}   \quad \quad \left ( \frac{   \lambda K  }{  H^{2}  } \leq 1  \right    )  , 
\label{f(H)_approx_C}
\end{equation}
where $H_{0}$ is the Hubble parameter at the present time and $\lambda K / (2 H_{0}^{2})$ is a dimensionless constant (corresponding to $\alpha_{4}$ in Ref.\ \cite{Koma6}). Hereafter, we call the constant-like term the constant term. Under the specific condition given by Eq.\ (\ref{f(H)_approx_C}), the Friedmann, acceleration, and continuity equations are written as
\begin{equation}
    H^2     =  \frac{ 8\pi G }{ 3 } \rho  +  \left ( \frac{\lambda K}{2H_{0}^{2}}  \right )  H_{0}^{2}  ,  
\label{N-N_FRW01_SR1_CST}
\end{equation}
\begin{equation}
  \frac{ \ddot{a} }{ a }    =   -  \frac{ 4 \pi G }{ 3} \left (  \rho  + \frac{3p}{c^{2}}  \right ) + \left ( \frac{\lambda K}{2H_{0}^{2}}  \right )  H_{0}^{2}  ,         
\label{N-N_FRW02_SR1_CST}
\end{equation}
\begin{equation}
       \dot{\rho} + 3  \frac{\dot{a}}{a} \left (  \rho + \frac{p}{c^2}  \right )    =  0       .
\label{drho_SR1_CST}
\end{equation}
The three equations are essentially the same as those in the standard $\Lambda$CDM model. Setting $p=0$ for non-relativistic matter, the background evolution of the universe is analytically given by 
\begin{align}
 \left ( \frac{H} {H_{0}} \right )^{2}   
&=  \left ( 1-   \frac{\lambda K}{2H_{0}^{2}}   \right )   \left ( \frac{ a } {  a_{0} } \right )^{-3}   +   \frac{\lambda K}{2H_{0}^{2}}      \notag \\ 
&=  \left ( 1-   \Omega_{h}  \right )   \left ( \frac{ a } {  a_{0} } \right )^{-3}   +   \Omega_{h}     , 
\label{eq:H/H0_Cst_A}
\end{align}
where $a_{0}$ represents the scale factor at the present time and $\Omega_{h}$ is a holographic parameter defined by
\begin{equation}
                   \Omega_{h}        \equiv  \frac{\lambda K}{ 2 H_{0}^{2} }   .                                             
\label{Omega-h}
\end{equation}
Using Eq.\ (\ref{Omega-h}) and $\tilde{H} \equiv H/H_{0}$, Eq.\ (\ref{eq:X-def}) can be written as
\begin{equation}
 X \equiv \frac{   \lambda K  }{  H^{2}  } = \frac{  \frac{2 \lambda K }{ 2 H_{0}^{2} }   }{   H^{2} /H_{0}^{2}  } = \frac{ 2 \Omega_{h}  }{   \tilde{H}^{2}  }    \quad \quad ( > 0) .  
\label{LK_OH_1}
\end{equation}
In the present model, the constant term $\lambda K /2$ and the holographic parameter $\Omega_{h}$ have been restricted by the inequality given in Eq.\ (\ref{f(H)_approx_C}). Accordingly, the constraint on both $\lambda K /2$ and $\Omega_{h}$ can be discussed, without tuning.

In the history of an expanding universe, $H_{0}$ is expected to be the minimum value of $H$, because $r_{H}$ increases with time and $H$ is given by $H = c/r_{H}$ from Eq.\ (\ref{eq:rH}).
Therefore, when $H= H_{0}$, the most severe constraint on $\lambda K$ can be obtained from the inequality of Eq.\ (\ref{f(H)_approx_C}).
The constraint is written as
\begin{equation}
  \lambda K  \lessapprox  H_{0}^{2}         .
\label{Cond_1}
\end{equation}
In the present model, Eq.\ (\ref{Cond_1}) is mathematically required to obtain the constant term.
From Eq.\ (\ref{Cond_1}), the order of the constant term, $\lambda K /2$, is approximately given by
\begin{equation}
  O \left ( \frac{\lambda K}{2} \right )  \lessapprox     O \left (  H_{0}^{2}   \right )       ,
\label{Cond_2}
\end{equation}
or equivalently, from Eqs.\ (\ref{Omega-h}) and (\ref{Cond_1}),  the order of $\Omega_{h}$ can be approximately written as
\begin{equation}
  O \left ( \Omega_{h} \right ) =  O \left ( \frac{\lambda K}{2 H_{0}^{2} } \right )  \lessapprox     O \left (  1   \right )       .
\label{Cond_2_2}
\end{equation}

We now compare the constant term $\lambda K /2$ in the present model and the cosmological constant term $\Lambda /3$ in the standard $\Lambda$CDM model. In the $\Lambda$CDM model, the density parameter $\Omega_{\Lambda}$ for the cosmological constant $\Lambda$ is given by 
 \begin{equation}
    \Omega_{\Lambda} \equiv  \frac{\Lambda}{3 H_{0}^{2}}  .
\label{Omega_L}
\end{equation}
Numerous cosmological observations indicate that the order of $\Omega_{\Lambda}$ is expected to be $1$, e.g., $\Omega_{\Lambda} =0.692$ from the Planck 2015 results \cite{Planck2015}.
Accordingly, the order of $\Omega_{\Lambda}$ can be approximately written as
 \begin{equation}
   O \left ( \Omega_{\Lambda}   \right )  \approx   O  \left (  1  \right ) . 
\label{L_order_0}
\end{equation} 
Using Eqs.\ (\ref{Omega_L}) and (\ref{L_order_0}), we have 
 \begin{equation}
   O \left ( \frac{\Lambda}{3 }  \right )  =  O \left ( \Omega_{\Lambda}  H_{0}^{2} \right )  \approx   O  \left (  H_{0}^{2}  \right ) . 
\label{L_order}
\end{equation}
Equations\ (\ref{Cond_2}) and (\ref{L_order}) imply that the order of the constant term $\lambda K /2$ is consistent with the order of $\Lambda /3$. That is, interestingly, the constant term in the present model is naturally consistent with the order of $\Lambda$ measured by cosmological observations as if it is $\Lambda$. Similarly, from Eqs.\ (\ref{Cond_2_2}) and (\ref{L_order_0}), the order of $\Omega_{h}$ is likely consistent with the order of $\Omega_{\Lambda}$. Of course, in the present model, $\lambda K  \lessapprox   H_{0}^{2}$ is required to obtain the constant term, assuming the modified R\'{e}nyi entropy instead of the Bekenstein--Hawking entropy. This extension may be beyond the original holographic equipartition law. 
However, the constant term and the specific condition considered here may provide new insights into the cosmological constant problem.

Equation\ (\ref{Cond_1}) can be written as $\lambda  \lessapprox  L_{p}^{2}/(\pi k_{B} r_{H0}^{2})  $ using Eq.\ (\ref{eq:K-def_3}) and $H_{0} = c/r_{H0}$, where $r_{H0}$ is the Hubble horizon at the present time.
This constraint indicates that $\lambda$ is an extremely small positive value.
In other words, the modified R\'{e}nyi entropy is approximately equivalent to the Bekenstein--Hawking entropy although it slightly deviates from the Bekenstein--Hawking entropy.
Accordingly, the present model may imply that the cosmological constant is related to a small deviation from the Bekenstein--Hawking entropy. As time passes, the constraint is expected to be more severe because $r_{H0}$ increases with time. Consequently, the extra driving term in the present model should gradually deviate from the constant value, even if the severe condition is satisfied at the present time. That is, the present model can be distinguished from the $\Lambda$CDM model, as discussed in Appendix \ref{Solutions for the present model and evolutions of the universe}.

In this paper, we have focused on the derivation of an extra driving term from the holographic equipartition law with the modified R\'{e}nyi entropy. The obtained term $ f(H)$ given by Eq.\ (\ref{f(H)_2}) is a logarithmic formula. However, under two specific conditions, it can be systematically written as $ f(H) = C_{0} H_{0}^{2}  + C_{1} H^{2}$, where $C_{0}$ and $C_{1}$ are dimensionless constants. In the $\Lambda(t)$CDM model, the above formula has been examined in detail \cite{Sola_2016_1,Valent2015,LimaSola_2013a,Basilakos1-Sola_2014a}.
For example, G\'{o}mez-Valent \textit{et al.} \cite{Valent2015} and Basilakos \textit{et al.} \cite{Basilakos1-Sola_2014a} have shown that it is not the $H^{2}$, $\dot{H}$, and $H$ terms but rather the constant term that plays an important role in cosmological fluctuations related to structure formations \cite{Koma9}. In addition, recently, $\Lambda (t)$CDM models, which include power series of $H$, have been found to be more suitable than the standard $\Lambda$CDM model \cite{Sola_2015L14}. In particular, the simple combination of the constant and $H^{2}$ terms, $ f(H) = C_{0} H_{0}^{2} + C_{1} H^{2}$, is likely favored; see  e.g., the works of Sol\`{a} \textit{et al.} \cite{Sola_2016_1},  G\'{o}mez-Valent \textit{et al.} \cite{Valent2015}, and Lima \textit{et al.} \cite{LimaSola_2013a}, which indicate that $C_{0}$ is dominant and $C_{1}$ is expected to be small \cite{Sola_2015L14,Sola_2016_1}. The smallness of $C_{1}$ has not yet been explained by the holographic approach \cite{Koma9},
though it may be explained by a deeper understanding of the present model. For example, the logarithmic formula discussed in the present model slightly deviates from the constant value when $\lambda K /H^{2} \approx 1$, as shown in Fig.\ \ref{Fig-f-X}. Of course, this small deviation is not proportional to $H^{2}$. However, the small deviation may be related to the smallness of $C_{1}$, if the smallness can be interpreted as a deviation from a constant term. This task is left for future research.

Keep in mind that the present result depends on the choice of entropy. 
In this paper, the modified R\'{e}nyi entropy \cite{Czinner1,Czinner2} is employed, in which the Bekenstein-Hawking entropy is considered to be a nonextensive Tsallis entropy.
That is, the Bekenstein-Hawking entropy is assumed to satisfy the Tsallis composition rule.
In addition, a logarithmic formula of the original R\'{e}nyi entropy is used for the modified R\'{e}nyi entropy \cite{Czinner1,Czinner2}, where the logarithmic formula is obtained from the Tsallis composition rule. 
For details of the modified R\'{e}nyi entropy, see Refs.\ \cite{Czinner1,Czinner2}.
The physical origin of the modified R\'{e}nyi entropy is likely unclear at the present time. 
However, it is worthwhile to examine cosmological models based on the holographic equipartition law from various viewpoints.
Note that the choice of temperature does not affect main results in the present study, because the extra driving term discussed here is related to $N_{\rm{sur}} = 4 S_{H} / k_{B}$ from Eq.\ (\ref{N_sur}) and is independent of the temperature \cite{C1}.

\section{Conclusions}
\label{Conclusions}

Recently, a novel type of R\'{e}nyi entropy on a black-hole horizon was proposed by Bir\'{o} and Czinner \cite{Czinner1}, in which a logarithmic formula of the original R\'{e}nyi entropy was used and the Bekenstein--Hawking entropy was regarded as a nonextensive Tsallis entropy. 
In this study, the modified R\'{e}nyi entropy has been applied to the holographic equipartition law proposed by Padmanabhan \cite{Padma2012A}, to investigate cosmological models based on the holographic equipartition law. 
Consequently, the acceleration equation, which includes an extra driving term, can be derived in a homogeneous, isotropic, and spatially flat universe. The extra driving term is a logarithmic formula which can explain the accelerated expansion of the late universe, as for effective dark energy.

Under two specific conditions, the extra driving term is found to be the constant-like and $H^{2}$-like terms, respectively. In particular, when a specific condition is satisfied, the extra driving term is found to be constant-like, i.e., $\lambda K /2$, as if it is a cosmological constant $\Lambda$. In the present model, $\lambda K  \lessapprox  H_{0}^{2}$ is required to obtain the constant-like term from the most severe constraint. In other words, the constant-like term must be extremely small when the extra driving term behaves as if it is constant. The present model may imply that the cosmological constant is related to a small deviation from the Bekenstein--Hawking entropy. In addition, interestingly, the order of the constant-like term is naturally consistent with the order of the cosmological constant measured by observations, because the specific condition constrains the value of the constant-like term. 
The present model may provide new insights into the cosmological constant problem.

This paper focuses on the derivation of the extra driving term and the specific condition for the constant-like term. Accordingly, general solutions for the present model that includes a logarithmic driving term have been separately studied in Appendix \ref{Solutions for the present model and evolutions of the universe}. To solve the cosmological equations, the present model is assumed to be a particular case of $\Lambda(t)$CDM models. From the obtained solution, the background evolutions of the universe in the present model are found to agree well with those in the standard $\Lambda$CDM model when both $2 \Omega_{h}/ \tilde{H}^2$ and $2 \Omega_{h}$ are small. For lower redshift, the present model gradually deviates from the $\Lambda$CDM model, due to the logarithmic term. Therefore, the present model can be distinguished from the standard $\Lambda$CDM model.

\appendix

\section{General Solutions for the present model and background evolutions of the late universe} 
\label{Solutions for the present model and evolutions of the universe}

So far, a constant term under a specific condition in the present model has been focused on, without tuning $\Omega_{h}$ given by Eq.\ (\ref{Omega-h}). However, in general, the present model has a logarithmic driving term. Accordingly, in this Appendix, general solutions for the present model are examined, assuming that it is a particular case of $\Lambda(t)$CDM models. Using the solution, background evolutions of the late universe are briefly observed. In the following, $p$ is set to be zero for non-relativistic matter. In addition, $\Omega_{h}$ is considered to be a free constant parameter (i.e., a kind of density parameter for effective dark energy), as for $\Omega_{\Lambda}$ in the standard $\Lambda$CDM model.

\subsection{General solutions for the present model} 
\label{Solutions}

In this subsection, the general solution for the present model is examined. To this end, the present model is assumed to be a particular case of $\Lambda(t)$CDM models, as shown in Eqs.\ (\ref{N-N_FRW01_SR1})--(\ref{drho_SR1}). From Eq.\ (\ref{f(H)_2}), the extra driving term $f(H)$ is written as
\begin{equation}
 f(H)  = H^{2}  \left (  1 - \frac{  \ln [1+ (\lambda K /H^{2}) ]  }{  \lambda K /H^{2} }  \right )  .
\label{f(H)_2_A}
\end{equation}
When $f(H) = \lambda K/2$, the solution is given by Eq.\ (\ref{eq:H/H0_Cst_A}). The solution method for the present model is partially based on Refs.\ \cite{Koma4,Koma5,Koma6}. (When $f(H)$ is a power series of $H$ and $\dot{H}$, see e.g., the recent works of G\'{o}mez-Valent \textit{et al.} \cite{Valent2015} and Basilakos \textit{et al.} \cite{Basilakos1-Sola_2014a}. )

Using $\ddot{a}/{a} = \dot{H}+ H^{2}$, coupling Eq.\ (\ref{N-N_FRW01_SR1}) with Eq.\ (\ref{N-N_FRW02_SR1})  $ \times 2$, and setting $p=0$, we obtain
\begin{equation}
    \dot{H} + \frac{3}{2} H^{2}  - \frac{3}{2} f(H)      =  0 ,  
\label{Back1}
\end{equation}
or, equivalently, 
\begin{equation}
    \dot{H} = - \frac{3}{2} H^{2}  \left ( 1 -  \frac{f(H)}{H^{2}}  \right )     .  
\label{Back2}
\end{equation}
From Eq.\ (\ref{Back2}), we have $(dH/da) a$ given by 
\begin{align}
\left ( \frac{dH}{da} \right )  a     &=       \left ( \frac{dH}{dt} \right )   \frac{dt}{da} a  
                                                   = - \frac{3}{2} H^{2}  \left ( 1 -  \frac{f(H)}{H^{2}}  \right )    \frac{a}{\dot{a}}                      \notag \\
                                                 &= - \frac{3}{2} H^{2}  \left ( 1 -  \frac{f(H)}{H^{2}}  \right )    \frac{1}{H}                              \notag \\     
                                                 &= - \frac{3}{2} H  \left ( 1 -  \frac{f(H)}{H^{2}}  \right )                                                      .          
\label{Back23}
\end{align}
Substituting Eq.\ (\ref{f(H)_2_A}) into Eq.\ (\ref{Back23}) gives 
\begin{align}
 \left ( \frac{dH}{da} \right )  a  &=  - \frac{3}{2} H  \left ( 1 -  \frac{H^{2}  \left (  1 - \frac{  \ln [1+ (\lambda K /H^{2}) ]  }{  \lambda K /H^{2} }  \right )  }{H^{2}}  \right )        \notag \\ 
                                               &=  - \frac{3}{2} H  \left (  \frac{  \ln [1+ (\lambda K /H^{2}) ]  }{  \lambda K /H^{2}    }  \right )                                                                     \notag \\
                                               &=  - \frac{3}{2} H  \left (  \frac{  \ln [1+ ( 2 \Omega_{h} /\tilde{H}^{2}) ]  }{ 2 \Omega_{h} /\tilde{H}^{2}    }  \right )                                  ,                                  
\label{Back24_f}
\end{align}
where the following equation given by Eq.\ (\ref{LK_OH_1}) has been also used:
\begin{equation}
                          \frac{   \lambda K  }{  H^{2}  }  = \frac{ 2 \Omega_{h}  }{   \tilde{H}^{2}  }  . 
\label{LK_OH_2}
\end{equation}
Note that, from Eq.\ (\ref{Omega-h}), $\Omega_{h}$ is defined by  
\begin{equation}
 \Omega_{h}  \equiv  \frac{ \lambda K }{ 2 H_{0}^{2} }  , 
\label{Omega-h2}
\end{equation}
and the normalized Hubble parameter $\tilde{H}$ is defined as
\begin{equation}
 \tilde{H} \equiv \frac{H}{H_{0}}  .
\label{def_H_H0}
\end{equation}
Similarly, the normalized scale factor $\tilde{a}$ is defined as
\begin{equation}
 \tilde{a} \equiv \frac{a}{a_{0}} . 
\label{def_a_a0}
\end{equation}
Substituting  $ H = \tilde{H} H_{0} $ and $a = \tilde{a} a_{0} $ into Eq.\ (\ref{Back24_f}), and arranging the resultant equation, we have 
\begin{equation}
 \left ( \frac{d \tilde{H}  }{d \tilde{a}  } \right )  \tilde{a}     =  - \frac{3}{2} \tilde{H}  \left (  \frac{  \ln [1+ ( 2 \Omega_{h} /\tilde{H}^{2}) ]  }{ 2 \Omega_{h} /\tilde{H}^{2}    }  \right )             .
\label{Back24_f_2}
\end{equation}
In addition, a parameter $N$ is defined by 
\begin{equation}
   N  \equiv \ln \tilde{a}, \quad \textrm{and therefore}  \quad   dN   = \frac{ d \tilde{a} }{ \tilde{a} }  .
\label{eq:N}
\end{equation}
Using Eq.\ (\ref{eq:N}), Eq.\ (\ref{Back24_f_2}) can be written as 
\begin{equation}
    \left ( \frac{d \tilde{H}  }{d N } \right )      =   - \frac{3}{2} \tilde{H}  \left (  \frac{  \ln [1+ ( 2 \Omega_{h} /\tilde{H}^{2}) ]  }{ 2 \Omega_{h} /\tilde{H}^{2}    }  \right )            .  
\label{Back24_f_N}
\end{equation}

When $\Omega_{h}$ is constant, Eq.\ (\ref{Back24_f_N}) can be integrated as 
\begin{equation}
    \int \frac{d \tilde{H}  }{\frac{3}{2} \tilde{H}  \left (  \frac{  \ln [1+ ( 2 \Omega_{h} /\tilde{H}^{2}) ]  }{ 2 \Omega_{h} /\tilde{H}^{2}    }  \right )   }       =   -   \int dN         .        
\label{f_N_I}
\end{equation}
This solution is given by
\begin{equation}
  - \frac{1}{3} \textrm{li}  \left (  1+  \frac{2 \Omega_{h} }{\tilde{H}^{2}}   \right )    =   -   N  + C = - \ln \tilde{a} + C        , 
\label{f_N_Solve}
\end{equation}
where $C$ is an integral constant and `li' represents the logarithmic integral \cite{LogInt} defined as 
\begin{equation}
 \textrm{li}(x) \equiv  \int^{x}_{0}  \frac{ dt }{ \ln t } .
\label{li}
\end{equation}
From Eqs.\ (\ref{def_H_H0}) and (\ref{def_a_a0}), the present values of $\tilde{H}$ and $\tilde{a}$ are $1$.
Substituting $\tilde{H} =1$ and $\tilde{a} =1$ into Eq.\ (\ref{f_N_Solve}), the integral constant $C$ can be written as 
\begin{equation}
 C = - \frac{1}{3} \textrm{li}  \left (  1+  2 \Omega_{h}   \right )        . 
\label{IntC}
\end{equation}
Substituting Eq.\ (\ref{IntC}) into Eq.\ (\ref{f_N_Solve}), and solving the resultant equation with respect to $\tilde{a}$, we obtain 
\begin{equation}
 \tilde{a} = \left \{ \exp \left [  \textrm{li}  \left (  1+  \frac{2 \Omega_{h} }{\tilde{H}^{2}}   \right )   -  \textrm{li}  \left (  1+  2 \Omega_{h}   \right  )     \right ] \right \}^{\frac{1}{3}}    . 
\label{Solution}
\end{equation}
This equation is the general solution for the present model. 
The relationship between $\tilde{a}$ and $\tilde{H}$ for each $\Omega_{h}$ can be calculated from Eq.\ (\ref{Solution}). 
Equations\ (\ref{Solution}) and (\ref{eq:H/H0_Cst_A}) approach the same approximate equation, respectively, when both $ 2 \Omega_{h} /\tilde{H}^{2} \ll 1 $ and $2 \Omega_{h} \ll 1$ and when both $\Omega_{\Lambda} /\tilde{H}^{2} \ll 1 $ and $\Omega_{\Lambda} \ll 1$.
The two conditions, $2 \Omega_{h} \ll 1$ and $\Omega_{\Lambda} \ll 1$, are related to the integral constants.

\subsection{Background evolutions of the late universe} 
\label{Evolutions of the universe in the present model}

In this subsection, the background evolutions of the late universe in the present model are briefly examined. For this purpose, $\Omega_{h}$ is considered to be a free parameter. Note that, when Eq.\ (\ref{Solution}) for $\Omega_{h}= 0$ is calculated numerically, $\Omega_{h}$ is set to be $10^{-4}$.

To observe the properties of the present model, the luminosity distance is examined. The luminosity distance \cite{Sato1} is generally given by 
\begin{equation}
  \left ( \frac{ H_{0} }{ c } \right )   d_{L}  
      =   (1+z)  \int_{1}^{1+z}  \frac{dy} { F(y) }    , 
\label{eq:dL_00}  
\end{equation}
where the integrating variable $y$, the function $F(y)$, and the redshift $z$ are given by 
\begin{equation}
  y = \tilde{a}^{-1},  \hspace{2mm} F(y)  = \tilde{H}, \hspace{2mm}    z \equiv \tilde{a}^{-1} -1  . 
\label{yFz}  
\end{equation}
The relationship between $\tilde{a}$ and $\tilde{H}$ for each $\Omega_{h}$ is obtained from Eq.\ (\ref{Solution}). Using this relationship, the luminosity distance is calculated from Eqs.\ (\ref{eq:dL_00}) and (\ref{yFz}).

\begin{figure} [t]  
\begin{minipage}{0.495\textwidth}
\begin{center}
\scalebox{0.3}{\includegraphics{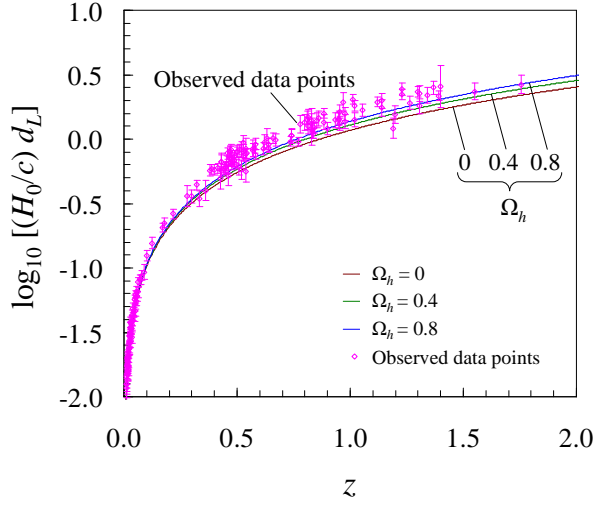}}
\end{center}
\end{minipage}
\caption{ (Color online). Dependence of the luminosity distance $d_L$ on redshift $z$. The solid lines represent $d_L$ for the present model. The open diamonds with error bars are supernova data points taken from Ref.\ \cite{Riess2007SN1}. For the supernova data points, $H_{0}$ is set to $67.8$ km/s/Mpc from the Planck 2015 results \cite{Planck2015}. }
\label{Fig-dL-z}
\end{figure}

For typical results, the luminosity distances for $\Omega_{h}= 0$, $0.4$, and $0.8$ are plotted in Fig.\ \ref{Fig-dL-z}. The luminosity distance for the present model tends to be more consistent with supernova data points with increasing $\Omega_{h}$. This result indicates that the present model can describe an accelerated expansion of the late universe. However, the influence of the increase in $\Omega_{h}$ on $d_{L}$ is likely to be smaller than the influence of the increase in $\Omega_{\Lambda}$ in the $\Lambda$CDM model on $d_{L}$ (the result for the $\Lambda$CDM model is not shown). To compare the two models, a temporal deceleration parameter is examined.

The temporal deceleration parameter $q$ is defined by 
\begin{equation}
q \equiv  - \left ( \frac{\ddot{a} } {a H^{2}} \right )  , 
\label{eq:q_def}
\end{equation}
where positive $q$ represents deceleration and negative $q$ represents acceleration. It should be noted that the $q$ examined here is not the nonextensive parameter discussed in previous sections. Substituting Eq.\ (\ref{Back2}) into $\ddot{a}/a = \dot{H} + H^{2}$, dividing the resultant equation by $- H^{2}$, and applying Eq.\ (\ref{eq:q_def}), we have
\begin{equation}
 q=        \frac{1}{2}   -  \frac{3}{2} \frac{ f(H) }{ H^{2} }   .
\label{eq:q_f(H)}
\end{equation}
In addition, substituting Eq. (\ref{f(H)_2_A}) into Eq.\ (\ref{eq:q_f(H)}), and applying Eq.\ (\ref{LK_OH_2}), we obtain 
\begin{align}
 q  &=    \frac{1}{2}   -  \frac{3}{2} \frac{ f(H) }{ H^{2} } = \frac{1}{2}   -  \frac{3}{2} \frac{ H^{2}  \left (  1 - \frac{  \ln [1+ (\lambda K /H^{2}) ]  }{  \lambda K /H^{2} }  \right )  }{ H^{2} }     \notag \\
     &=    \frac{1}{2}   -  \frac{3}{2}  \left (  1 - \frac{  \ln [1+ (\lambda K /H^{2}) ]  }{  \lambda K /H^{2} }  \right )       \notag \\
     &=       -1           +  \frac{3}{2}  \left (  \frac{  \ln [1+ ( 2 \Omega_{h} /\tilde{H}^{2}) ]  }{ 2 \Omega_{h} /\tilde{H}^{2}    }  \right )       .
\label{eq:q_Present1}
\end{align}
The temporal deceleration parameter $q$ for the present model can be calculated from Eq.\ (\ref{eq:q_Present1}). Note that the relationship between $\tilde{a}$ and $\tilde{H}$ is obtained from Eq.\ (\ref{Solution}).

For the $\Lambda$CDM model, substituting $f(H) = \Lambda / 3$ into Eq.\ (\ref{eq:q_f(H)}), and using $\Omega_{\Lambda} = \Lambda /( 3 H_{0}^{2} )$ and $\tilde{H}=H/H_{0}$, we have
\begin{equation}
 q=        \frac{1}{2}   -  \frac{3}{2} \frac{ \Omega_{\Lambda} }{ \tilde{H}^{2} }   , 
\label{eq:q_LCDM}
\end{equation}
where  $\tilde{H}^{2}$ is given by $\tilde{H}^{2} =  ( 1-   \Omega_{\Lambda} ) \tilde{a}^{-3}   +   \Omega_{\Lambda}$. In a flat universe, the density parameter $\Omega_{m}$ for matter is given by $\Omega_{m} = 1- \Omega_{\Lambda}$, neglecting the density parameter for the radiation.

\begin{figure} [t] 
\begin{minipage}{0.495\textwidth}
\begin{center}
\scalebox{0.3}{\includegraphics{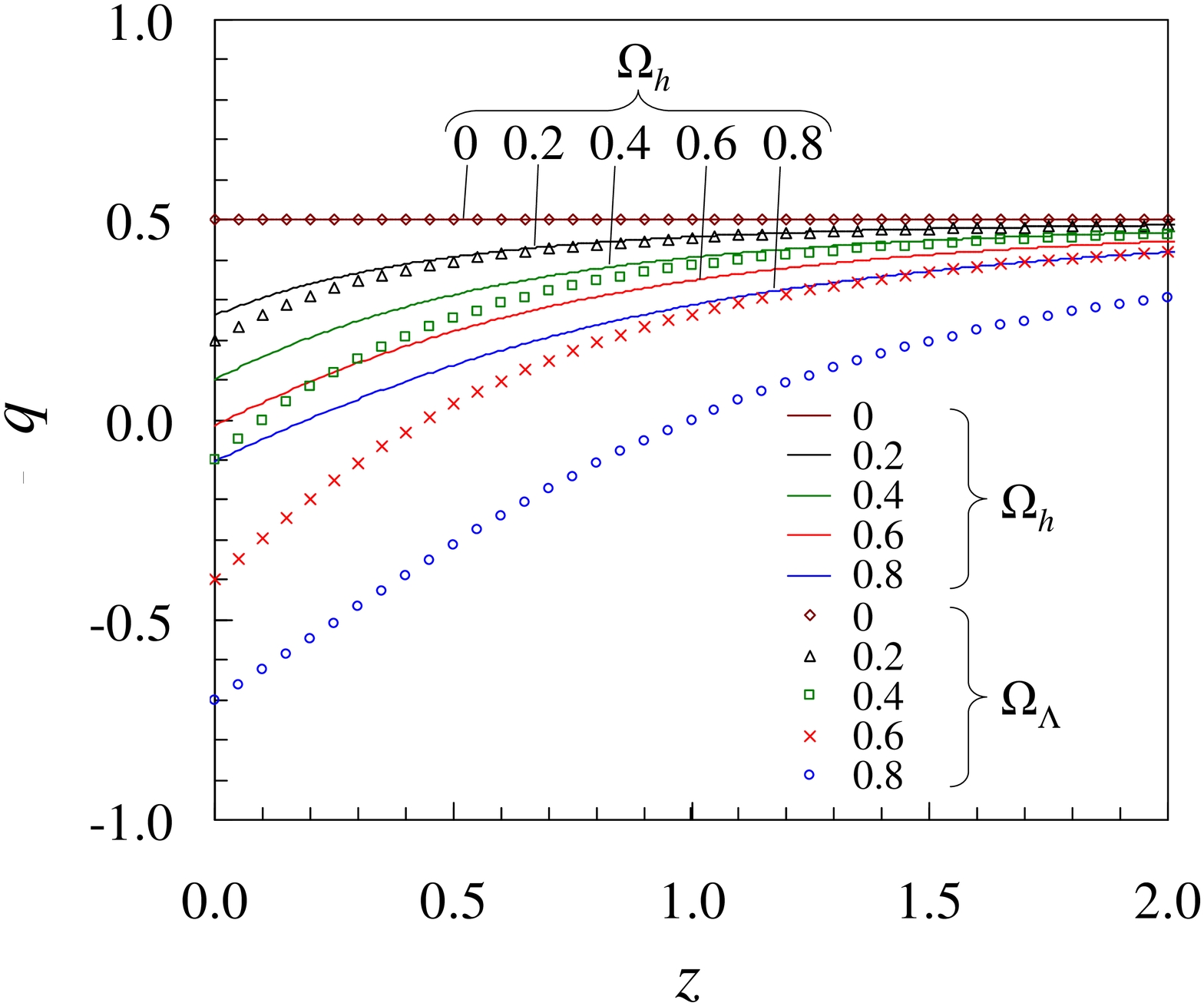}}
\end{center}
\end{minipage}
\caption{ (Color online). Dependence of the temporal deceleration parameter $q$ on redshift $z$. The solid lines represent $q$ for the present model, while the symbols represent $q$ for the $\Lambda$CDM model.
Note that positive $q$ represents deceleration and negative $q$ represents acceleration. }
\label{Fig-q-z}
\end{figure}

\begin{figure} [t] 
\begin{minipage}{0.495\textwidth}
\begin{center}
\scalebox{0.3}{\includegraphics{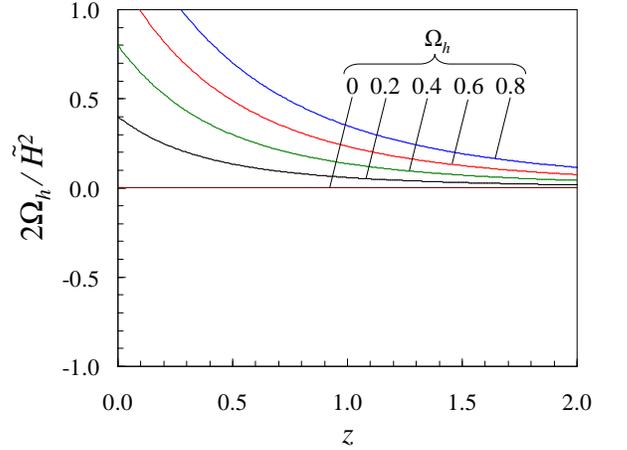}}
\end{center}
\end{minipage}
\caption{ (Color online). Dependence of the parameter $2\Omega_{h} / \tilde{H}^{2}$ on redshift $z$.
Note that $2 \Omega_{h}/ \tilde{H}^2$ is equivalent to $X \equiv \lambda K /H^{2}$ shown in Fig. \ref{Fig-f-X}. }
\label{Fig-OH-z}
\end{figure}
Figure\ \ref{Fig-q-z} shows the dependence of the temporal deceleration parameter $q$ on the redshift $z$. As expected, $q$ for $\Omega_{h}= 0$ and $0.2$ is consistent with that for $\Omega_{\Lambda}= 0$ and $0.2$, respectively. However, $q$ for the present model gradually deviates from that for the $\Lambda$CDM model, with increasing $\Omega_{h}$, especially for low redshift. In fact, even for $\Omega_{h}= 0.2$, $q$ gradually deviates from that for $\Omega_{\Lambda}= 0.2$ with decreasing $z$, although the two results agree well for high redshift. As shown in Fig.\ \ref{Fig-OH-z}, $2 \Omega_{h}/ \tilde{H}^2$ increases with decreasing $z$, because $\tilde{H}$ decreases with decreasing $z$. That is, the deviation from the $\Lambda$CDM model for low redshift (shown in Fig.\ \ref{Fig-q-z}) is related to the increase in $2 \Omega_{h}/ \tilde{H}^2$. In addition, for $\Omega_{h} =0.6$ and $0.8$, $q$ varies from positive to negative with decreasing $z$, as shown in Fig.\ \ref{Fig-q-z}. Therefore, the present model can describe a decelerated and accelerated expansion of the late universe.

In this Appendix, the general solution for the present model has been examined, assuming that the present model is a particular case of $\Lambda(t)$CDM models. In addition, background evolutions of the late universe in the present model have been discussed. Consequently, the present model is found to agree well with the standard $\Lambda$CDM model, when both $2 \Omega_{h}/ \tilde{H}^2$ and $2 \Omega_{h}$ are small. For lower redshift, the present model gradually deviates from the $\Lambda$CDM model, due to the properties of the logarithmic driving term. Accordingly, the present model examined here can be distinguished from the $\Lambda$CDM model.

\begin{acknowledgements}
The author wishes to thank H. Iguchi and V. G. Czinner for very valuable comments. 
\end{acknowledgements}


\begin{thebibliography}{99}


%
\bibitem{PERL1998_Riess1998} S. Perlmutter \textit{et al.},  Nature (London) \textbf{391}, 51 (1998); 
A. G. Riess \textit{et al.}, Astron. J. \textbf{116}, 1009 (1998).
\bibitem{Riess2007SN1} A. G. Riess \textit{et al.},     Astrophys. J. \textbf{659},   98 (2007); 
http://braeburn.pha.jhu.edu/\~{}ariess/R06/sn\_sample.


\bibitem{Planck2015} P. A. R. Ade \textit{et al.}, Astron. Astrophys. \textbf{594}, A13 (2016).     

\bibitem{Weinberg1989} S. Weinberg, Rev. Mod. Phys. \textbf{61}, 1 (1989);  I. Zlatev, L. Wang, P. J. Steinhardt, Phys. Rev. Lett.  \textbf{82}, 896 (1999); 
V. Sahni, A. A. Starobinsky, Int. J. Mod. Phys. D \textbf{9}, 373 (2000); 
S. M. Carroll, Living Rev. Relativity \textbf{4}, 1 (2001); 
T. Padmanabhan, Phys. Rep. \textbf{380}, 235 (2003);
%
J. D. Barrow, D. J. Shaw, Phys. Rev. Lett. \textbf{106}, 101302, (2011); 
S. Finazzi, S. Liberati, L. Sindoni, Phys. Rev. Lett. \textbf{108}, 071101, (2012);
%
J. Sol\`{a}, E. Karimkhani, A. Khodam-Mohammadi, arXiv:1609.00350.





\bibitem{Weinberg1} S. Weinberg, \textit{Cosmology} (Oxford University Press, New York, 2008).

\bibitem{Bamba1} K. Bamba, S. Capozziello, S. Nojiri, S. D. Odintsov, Astrophys. Space Sci. \textbf{342}, 155 (2012).









%
%
\bibitem{Prigogine_1988-1989}  
I. Prigogine, J. Geheniau, E. Gunzig, P. Nardone, Proc. Natl. Acad. Sci. U.S.A. \textbf{85}, 7428 (1988); 
Gen. Relativ. Gravit. \textbf{21}, 767 (1989).

\bibitem{Lima-Others1996-2016}   
J. A. S. Lima, A. S. M. Germano, L. R. W. Abramo, Phys. Rev. D \textbf{53}, 4287 (1996); 
W. Zimdahl, D. J. Schwarz, A. B. Balakin, D. Pav\'{o}n, Phys. Rev. D \textbf{64}, 063501 (2001); 
%
S. Basilakos, M. Plionis, J. A. S. Lima, Phys. Rev. D \textbf{82}, 083517 (2010);
J. A. S. Lima, S. Basilakos, F. E. M. Costa, Phys. Rev. D \textbf{86}, 103534 (2012);
%
J. F. Jesus, S. H. Pereira, J. Cosmol. Astropart. Phys. 07 (2014) 040; 
%
%
J. A. S. Lima, R. C. Santos, J. V. Cunha, J. Cosmol. Astropart. Phys. 03 (2016) 027.


%
\bibitem{Lima_1992-2016}   
M. O. Calv\~{a}o, J. A. S. Lima, I. Waga, Phys. Lett. A \textbf{162}, 223 (1992); 
%
J. A. S. Lima, A. I. Silva, S. M. Viegas, Mon. Not. R. Astron. Soc. \textbf{312}, 747 (2000); 
%
T. Harko, Phys. Rev. D \textbf{90}, 044067 (2014); 
%
I. Baranov, J. F. Jesus, J. A. S. Lima, arXiv:1605.04857. 


\bibitem{Freese-Mimoso_2015}
K. Freese, F. C. Adams, J. A. Frieman, E. Mottola, Nucl. Phys. \textbf{B287}, 797 (1987); 
J. M. Overduin, F. I. Cooperstock, Phys. Rev. D \textbf{58}, 043506 (1998); 
I. L. Shapiro, J. Sol\`{a}, J. High Energy Phys. 02 (2002) 006; 
H. Fritzsch, J. Sol\`{a}, Classical Quantum Gravity \textbf{29}, 215002, (2012); 
J. P. Mimoso, D. Pav\'{o}n, Phys. Rev. D \textbf{87}, 047302 (2013);
M. H. P. M. Putten, Mon. Not. R. Astron. Soc. \textbf{450}, L48 (2015).
%




\bibitem{Sola_2009-2015} S. Basilakos, M. Plionis, J. Sol\`{a}, Phys. Rev. D \textbf{80}, 083511 (2009);
J. Grande, J. Sol\`{a}, S. Basilakos, M. Plionis, J. Cosmol. Astropart. Phys. 08 (2011) 007; 
E. L. D. Perico, J. A. S. Lima, S. Basilakos, J. Sol\`{a}, Phys. Rev. D \textbf{88}, 063531 (2013); 
S. Basilakos, J. Sol\`{a}, Mon. Not. R. Astron. Soc. \textbf{437}, 3331 (2014); 
S. Basilakos, J. Sol\`{a}, Phys. Rev. D \textbf{92}, 123501 (2015).


\bibitem{Sola_2013_Review}
J. Sol\`{a}, J. Phys. Conf. Ser. \textbf{453},  012015 (2013).

\bibitem{Sola_2015_2}
J. Sol\`{a}, Int. J. Mod. Phys. D \textbf{24}, 1544027 (2015);
J. Sol\`{a}, A. G\'{o}mez-Valent, Int. J. Mod. Phys. D \textbf{24}, 1541003 (2015);
J. Sol\`{a}, Int. J. Mod. Phys. A \textbf{31}, 1630035 (2016).


\bibitem{Sola_2015L14}
J. Sol\`{a}, A. G\'{o}mez-Valent, J. C. P\'{e}rez, Astrophys. J. \textbf{811}, L14 (2015).
%
\bibitem{Sola_2016_1} 
J. Sol\`{a}, J. C. P\'{e}rez, A. G\'{o}mez-Valent, R. C. Nunes, arXiv.1606.00450.


\bibitem{Valent2015} 
A. G\'{o}mez-Valent, J. Sol\`{a}, Mon. Not. R. Astron. Soc. \textbf{448}, 2810 (2015);     
A. G\'{o}mez-Valent, J. Sol\`{a}, S. Basilakos, J. Cosmol. Astropart. Phys. 01 (2015) 004; 
A. G\'{o}mez-Valent, E. Karimkhani, J. Sol\`{a}, J. Cosmol. Astropart. Phys. 12 (2015) 048.

\bibitem{LimaSola_2013a} 
J. A. S. Lima, S. Basilakos, J. Sol\`{a},  Mon. Not. R. Astron. Soc. \textbf{431}, 923 (2013). 
%
\bibitem{LimaSola_2015-2016} 
J. A. S. Lima, S. Basilakos, J. Sol\`{a}, 
Gen. Relativ. Gravit. \textbf{47}, 40 (2015);         
Eur. Phys. J. C \textbf{76}, 228 (2016).
%
\bibitem{EPJC} 
M. Tong, H. Noh, Eur. Phys. J. C \textbf{71}, 1586 (2011);
Jing-Fei Zhang, Yang-Yang Li, Y. Liu, S. Zou, X. Zhang, Eur. Phys. J. C \textbf{72}, 2077 (2012); 
A. Stachowski, M. Szydlowski, Eur. Phys. J. C \textbf{76}, 606 (2016).


%
\bibitem{Bekenstein1}  J. D. Bekenstein, Phys. Rev. D \textbf{7}, 2333 (1973); Phys. Rev. D \textbf{9}, 3292 (1974);  Phys. Rev. D \textbf{12}, 3077 (1975).
\bibitem{Hawking1}  S. W. Hawking, Phys. Rev. Lett. \textbf{26}, 1344 (1971); Nature \textbf{248}, 30 (1974); Commun. Math. Phys. \textbf{43}, 199 (1975); Phys. Rev. D \textbf{13}, 191 (1976).
%

\bibitem{Hooft-Bousso}
G. 't Hooft, arXiv:gr-qc/9310026; L. Susskind, J. Math. Phys. \textbf{36}, 6377 (1995); R. Bousso, Rev. Mod. Phys. \textbf{74}, 825 (2002).

\bibitem{Jacob1995} T. Jacobson, Phys. Rev. Lett. \textbf{75}, 1260 (1995).
%
\bibitem{Padma1}  T. Padmanabhan, Mod. Phys. Lett. A \textbf{25}, 1129 (2010).
\bibitem{Verlinde1} E. Verlinde, J. High Energy Phys. 04 (2011) 029.
%
%
\bibitem{Padma2010}  T. Padmanabhan, Rept. Prog. Phys. \textbf{73}, 046901 (2010).
%
\bibitem{Sheykhi1}
A. Sheykhi, Phys. Rev. D \textbf{81}, 104011 (2010);
K. Karami, A. Sheykhi, N. Sahraei, S. Ghaffari, Eur. Phys. Lett. \textbf{93}, 29002 (2011); 
A. Sheykhi, S. H. Hendi, Phys. Rev. D \textbf{84}, 044023 (2011). 
%

\bibitem{Sadjadi1} H. M. Sadjadi, M. Jamil, Eur. Phys. Lett. \textbf{92}, 69001 (2010); 
%
S. Mitra, S. Saha, S. Chakraborty, Mod. Phys. Lett. A \textbf{30}, 1550058 (2015).



\bibitem{Easson12}  D. A. Easson, P. H. Frampton, G. F. Smoot, Phys. Lett. B \textbf{696}, 273 (2011); Int. J. Mod. Phys. A \textbf{27}, 1250066 (2012).
\bibitem{Koivisto-Costa1}  
Y. F. Cai, J. Liu, H. Li,      Phys. Lett. B \textbf{690}, 213 (2010); 
Y. F. Cai, E. N. Saridakis,  Phys. Lett. B \textbf{697}, 280 (2011); 
T. S. Koivisto, D. F. Mota, M. Zumalac\'{a}rregui, J. Cosmol. Astropart. Phys. 02 (2011) 027; 
T. Qiu, E. N. Saridakis,     Phys. Rev. D \textbf{85}, 043504 (2012).
%
\bibitem{Lepe1} S. Lepe, F. Pen\~{a}, arXiv:1201.5343v2 [hep-th].
\bibitem{Basilakos1-Sola_2014a}
S. Basilakos, D. Polarski, J. Sol\`{a}, Phys. Rev. D \textbf{86}, 043010 (2012); 
S. Basilakos, J. Sol\`{a}, Phys. Rev. D \textbf{90}, 023008 (2014).

\bibitem{Koma4}  N. Komatsu, S. Kimura, Phys. Rev. D \textbf{87}, 043531 (2013); N. Komatsu, JPS Conf. Proc. \textbf{1}, 013112 (2014).
\bibitem{Koma5}  N. Komatsu, S. Kimura, Phys. Rev. D \textbf{88}, 083534 (2013).
\bibitem{Koma6}  N. Komatsu, S. Kimura, Phys. Rev. D \textbf{89}, 123501 (2014).
\bibitem{Koma7}  N. Komatsu, S. Kimura, Phys. Rev. D \textbf{90}, 123516 (2014).
\bibitem{Koma8}  N. Komatsu, S. Kimura, Phys. Rev. D \textbf{92}, 043507 (2015).
\bibitem{Koma9}  N. Komatsu, S. Kimura, Phys. Rev. D \textbf{93}, 043530 (2016).
\bibitem{Gohar_2015ab} 
M. P. D\c{a}browski, H. Gohar, Phys. Lett. B \textbf{748}, 428 (2015); 
M. P. D\c{a}browski, H. Gohar, V. Salzano, Entropy \textbf{18}, 60 (2016).

\bibitem{Tsallis2012}  C. Tsallis, L. J. L. Cirto, Eur. Phys. J. C \textbf{73}, 2487 (2013). 

\bibitem{Padma2012A}  T. Padmanabhan, arXiv:1206.4916 [hep-th].
\bibitem{Padma2012}  T. Padmanabhan, Res. Astron. Astrophys. \textbf{12}, 891 (2012).
%
%
\bibitem{Cai2012-Tu2013}                                                                               
R. G. Cai, J. High Energy Phys. 1211 (2012) 016; 
K. Yang, Y.-X. Liu, Y.-Q. Wang, Phys. Rev. D \textbf{86} 104013 (2012); 
%
M. Eune, W. Kim, Phys. Rev. D \textbf{88} 067303 (2013); 
A. Sheykhi,  M. H. Dehghani, S. E. Hosseini, Phys. Lett. B \textbf{726}, 23 (2013); 
%
%
Fei-Quan Tu, Yi-Xin Chen, J. Cosmol. Astropart. Phys. 05 (2013) 024; 
%
%
A. F. Ali, Phys. Lett. B \textbf{732}, 335 (2014);
E. Chang-Young, D. Lee, J. High Energ. Phys. 04 (2014) 125; 
M. Hashemi, S. Jalalzadeh, S. Vasheghani Farahani, Gen. Relativ. Gravit. \textbf{47}, 53 (2015); 
%
F. L. Dezaki, B. Mirza, Gen. Relativ. Gravit. \textbf{47}, 67 (2015); 
%
%
H. Moradpour, Int. J. Theor. Phys. \textbf{55}, 4176 (2016);
%
S. Chakraborty, T. Padmanabhan, Phys. Rev. D \textbf{92}, 104011 (2015).


\bibitem{Padma2014-2015}  
T. Padmanabhan, H. Padmanabhan, Int. J. Mod. Phys. D \textbf{23}, 1430011 (2014); 
T. Padmanabhan, Mod. Phys. Lett. A \textbf{30}, 1540007 (2015).



\bibitem{ZLWang2015}  
Zi-Liang Wang, Wen-Yuan Ai, Hua Chen, Jian-Bo Deng, Phys. Rev. D \textbf{92}, 024051 (2015).
%
\bibitem{Tu2015}  
Fei-Quan Tu, Yi-Xin Chen, Gen. Relativ. Gravit. \textbf{47}, 87 (2015).


\bibitem{Ren1}    A. R\'{e}nyi, \textit{Probability Theory} (North-Holland, Amsterdam, 1970).
\bibitem{Tsa0}    C. Tsallis, J. Stat. Phys. \textbf{52},  479 (1988).
\bibitem{Tsa1}    C. Tsallis,  {\it Introduction to Nonextensive Statistical Mechanics: Approaching a Complex World} (Springer, New York, 2009).


\bibitem{Czinner1}       T. S. Bir\'{o}, V. G. Czinner, Phys. Lett. B \textbf{726}, 861 (2013).
\bibitem{Czinner2}       V. G. Czinner, H. Iguchi, Phys. Lett. B \textbf{752}, 306 (2016).


\bibitem{Plas1-Tsallis2001}    A. Plastino, A. R. Plastino, Phys. Lett. A \textbf{174}, 384 (1993); 
S. Abe, Phys. Lett. A \textbf{263}, 424 (1999); 
D. F. Torres, H. Vucetich, A. Plastino, Phys. Rev. Lett. \textbf{79}, 1588 (1997); 
R. S. Mendes, C. Tsallis, Phys. Lett. A \textbf{285}, 273 (2001).
\bibitem{Chava21-Liu}  
P. H. Chavanis, Astron. Astrophys. \textbf{386}, 732 (2002); 
A. Taruya, M. Sakagami,  Phys. Rev. Lett. \textbf{90},  181101 (2003); 
A. Nakamichi, M. Morikawa, Physica A \textbf{341}, 215 (2004); 
B. Liu, J. Goree,  Phys. Rev. Lett. \textbf{100}, 055003 (2008).
%
%
\bibitem{NonExtensive} 
P. T. Landsberg, D. Tranah, Phys. Lett. A \textbf{78}, 219 (1980); 
D. Pav\'{o}n, J. M. Rubi, Gen. Relativ. Gravit. \textbf{18}, 1245 (1986);
J. Oppenheim, Phys. Rev. E \textbf{68}, 016108 (2003); 
A. Bialas, W. Czyz, Europhys. Lett. \textbf{83}, 60009 (2008); 
A. Belin, A. Maloney, S. Matsuura, J. High Energy Phys. 12 (2013) 050.
%
\bibitem{Koma2-3}  N. Komatsu, S. Kimura, T. Kiwata, Phys. Rev. E \textbf{80}, 041107 (2009); N. Komatsu, T. Kiwata, S. Kimura, Phys. Rev. E \textbf{82}, 021118 (2010); Phys. Rev. E \textbf{85}, 021132 (2012).
%
\bibitem{Nunes_2015b}
E. M. Barboza Jr., R. C. Nunes, E. M. C. Abreu, J. A. Neto, Physica A \textbf{436}, 301 (2015);
R. C. Nunes, E. M. Barboza Jr., E. M. C. Abreu, J. A. Neto, arXiv:1509.05059v1.
%
%
\bibitem{Czinner2016}  V. G. Czinner, F. C. Mena, Phys. Lett. B \textbf{758}, 9 (2016).





%
\bibitem{Barrow22-2015}  
J. D. Barrow, T. Clifton, Phys. Rev. D \textbf{73}, 103520 (2006); 
S. Nojiri, S. D. Odintsov, Phys. Lett. B \textbf{639}, 144 (2006); 
%
Y. Wang, D. Wands, G.-B. Zhao, L. Xu, Phys. Rev. D \textbf{90}, 023502 (2014); 
N. Tamanini, Phys. Rev. D \textbf{92}, 043524 (2015).


\bibitem{C1}
A novel type of temperature on a black-hole horizon is proposed in Refs.\ \cite{Czinner1,Czinner2}.
However, an extra driving term examined in this study is related to $N_{\rm{sur}} = 4 S_{H} / k_{B}$ from Eq.\ (\ref{N_sur}) and is independent of the temperature, as shown in Sec.\ \ref{Holographic equipartition}. 
That is, the choice of temperature does not affect main results in the present study.
Therefore, in this paper, Eq.\ (\ref{eq:T0}) is used for the temperature on the Hubble horizon. 


\bibitem{LogInt}
M. Abramowitz, I. A. Stegun, \textit{Handbook of Mathematical Functions with Formulas, Graphs, and Mathematical Tables} (Dover, New York, 1972); 
http://mathworld.wolfram.com/LogarithmicIntegral.html.

\bibitem{Sato1}    K. Sato \textit{et al.}, \textit{Cosmology I}, Modern Astronomy Series Vol. 2, edited by K. Sato and T. Futamase (Nippon HyoronSha Co., Tokyo, 2008), in Japanese.



\end{thebibliography}
\end{document}